\documentclass[12pt,a4paper]{article}
\usepackage{epsfig}
\usepackage{amsfonts}
\usepackage{rotating}
\usepackage{cite}
\usepackage{float}

\newcommand{\beq}{\begin{equation}}
\newcommand{\eeq}{\end{equation}  }
\newcommand{\bec}{\begin{center}}
\newcommand{\eec}{\end{center}}
\def\d{\delta}

\def\as{\alpha_s}

\def\ran{\rangle}
\def\lan{\langle}
\newcommand{\PR}{Phys. Rev.\ }
\newcommand{\PRL}{Phys. Rev. Lett.\ }
\newcommand{\PL}{Phys. Lett.\ }
\newcommand{\NP}{Nucl. Phys.\ }
\newcommand{\ZP}{Z. Phys.\ }

\newcommand{\Coll}{Collaboration}
\newcommand{\CPC}{Comp. Phys. Comm. \ }

\makeatletter
   \newcommand\figcaption{\def\@captype{figure}\caption}
   \newcommand\tabcaption{\def\@captype{figure}\caption}
\makeatother

\begin{document}
\clearpage
\pagestyle{empty}
\setcounter{footnote}{0}\setcounter{page}{0}%
\thispagestyle{empty}\pagestyle{plain}\pagenumbering{arabic}%

\hfill ANL-HEP-PR-01-05 

\hfill April, 2001 

\vspace{2.0cm}

\begin{center}

{\Large\bf
Suppression of multi-gluon fluctuations\\  
in jets at a Linear Collider \\[-1cm]} 

\vspace{2.5cm}

{\large S.~Chekanov
\footnote[1]{On leave from
the Institute of Physics,  AS of Belarus,
Skaryna av.70, Minsk 220072, Belarus.}
}

Argonne National Laboratory,
HEP Division, 9700 S.Cass Avenue, \\ 
Argonne, IL 60439.
USA \\ Email: chekanov@mail.desy.de

\vspace{2.5cm}

\begin{abstract}
It was shown that at typical energies and 
luminosities  of an $e^+e^-$ linear collider 
one can observe a significant suppression of  
fluctuations of the number of gluons in small angular regions.
This property of jets is expected in  $e^+e^-$ annihilation as well as 
in  any collision  producing high energy quarks which give rise to   
the QCD gluon cascade.   
The reduction of multiplicity  fluctuations 
was obtained from perturbative QCD using bunching
parameters and then cross checked with  
a Monte Carlo simulation. The parton-level Monte Carlo simulation
confirms the prediction, while the hadron level illustrates 
that the effect survives the hadronization process.         
\end{abstract}

\end{center}
\newpage
\setcounter{page}{1}

\section{Introduction}

A hadronic jet produced  in high energy experiments can be viewed as
a spray of strongly correlated particles created by  a hard  
partonic scattering initiated at large momentum transfers 
($\gg \Lambda$ with $\Lambda$ being a characteristic QCD scale)
and subsequent parton cascade
followed by a soft fragmentation process.    
At high energies, the dominant source of
particle production inside jets 
is gluon splitting in the QCD cascade,
where the presence of  a gluon enhances the probability for
emission of another gluon nearby in momentum space.
This leads  to inter-parton
correlations and non-Poisson statistics for the
multiplicity distributions in the restricted  phase-space intervals
where partons are counted.

In the near future, experiments at 
the Large Hadron Collider (LHC) and, possibly,   
at an $e^+e^-$ linear collider (LC) will search for undiscovered Higgs bosons,
new physics beyond the Standard Model as well as will focus 
on the precise measurements of known electroweak gauge bosons.  
Such studies will heavily rely on our understanding of the QCD 
multiple gluon radiation which is presently simulated
with Monte Carlo (MC) models. These models
take into account higher than first-order QCD effects
in terms of the parton showers. This approach, however,  has some
shortcomings due to its approximate nature.  	 
Therefore, precise knowledge of 
gluon activity in jets together with 
reliable predictions of Monte Carlo  models will require
detailed and comprehensive studies of the jet structure.     

The internal structure of jets  can be investigated  using 
subjet multiplicities at some resolution scale $y_{cut}$,  
energy flows in specific kinematic variables with respect to the jet axis
or calculating the jet shapes \cite{el}. 
However, these studies do not address the issue of fluctuations
of the number of separate hadrons in jets on event-to-event basis
and do not resolve directly many-particle inclusive densities of the
hadronic final state.    
To measure such fluctuations, one can calculate the probability  
distribution $P_n(\d)$ of observing $n$ particles in a phase-space
interval of size $\d$  inside a jet. Note that, in this
case, we are interested   
not only in the average number $\lan n (\d )\ran$ 
of particles inside jets, but rather in the evolution
of the shape of $P_n(\d)$ as a function of the size $\d$. 
This can be studied using the  methods  to be discussed below.   
Such studies have been done recently by the LEP and HERA 
experiments using various tools  \cite{fluc_l3,ang_ex,ex} 
(see \cite{rev} for more references).

The semi-soft QCD physics, which determines  multiplicities
of the final-state hadrons inside jets,  will become one of the important
topics for investigation at  LHC and LC.  
With an increase of the center-of-mass (c.m.) energy of collisions, 
the multiplicities of 
produced hadrons inside jets will rise significantly. 
For example, for a LC with $\sqrt{s}=500$ GeV, the average number of hadrons
in jets will increase by a factor two with respect to the LEP1 experiments.
This would allow detailed studies of the jet structure by 
counting separate particles inside jets 
and establishing relations  between them in 
terms  of correlations/fluctuations. Such a rise of the 
average hadron  multiplicity, together with an increase in luminosity 
of the future
experiments,  will also allow the determination of the probability  
distribution $P_n(\d)$ for  multiplicities 
higher than at LEP, assuming that the size $\d$ will be 
close to that of the previous experiments.

\section{Methods}

One of the important problems in studying  of multiparticle systems is 
to develop  methods  which are sufficiently sensitive to the 
fluctuations of hadrons inside jets and    
make it possible an adequate comparison of theoretical predictions
with experimental data. 
When fluctuations of separate particles produced inside jets are measured, 
it is convenient to transform the multiplicity distribution
$P_n(\d )$ to  the following observables: 
\begin{eqnarray}
\mathrm{NFM:} &  &  
\quad F_q(\d  ) = \lan  n (\d  )\ran^{-q}  
\sum^{\infty}_{n=q}  \frac{n!}{(n-q)!} P_n(\d  ), \\ 
\mathrm{CFM:} &  & \quad K_q(\d ) = 
F_q(\d ) - \sum^{q-1}_{m=1} \frac{(q-1)!}{m!(q-m-1)!} 
K_{q-m}(\d ) F_m(\d ) , \\ 
\mathrm{BP:} &  & \quad \eta_q(\d  )=
\frac{q}{q-1}\frac{P_q(\d  )P_{q-2}(\d  )}
{P_{q-1}^2(\d  ) } , 
\end{eqnarray}
where the abbreviations  denote the normalized 
factorial moments (NFM) \cite{bi},
cumulant factorial moments (CFM) \cite{mu} and 
bunching parameters (BP) \cite{bp}. Note that
the experimental definitions of the NFM and BP given in
\cite{bi, bp} do not involve direct measurements of the $P_n(\d )$
and thus they differ from the theoretical formulae  above.  
These three quantities measure deviations of 
the multiplicity distribution $P_n(\d )$  
from a Poisson  distribution $P_n^{Posson}(\d )$, since
for this distribution
$F_q(\d )=\eta_q(\d  )=1$, $K_q(\d )=0$.
Note that such deviations are measured differently by these
three methods \cite{rev}.

Uncorrelated particle production inside $\d$ leads
to the Poisson statistics, thus 
deviations of the NFM, CFM and BP from the Poisson values 
mean correlations between particles and  dynamical fluctuations.
Let us note that, in a general case,   
uncorrelated phase-space production 
can have non-Poisson event-to-event
fluctuations. In this case the NFM, CFM and BP 
are $\d$-independent constants which can differ from unity.  

In this paper we study the $P_n(\d )$ at asymptotically
small $\d$, 
\beq
\lan  n (\d  )\ran \ll  1, 
\qquad P_n(\d  )\gg P_{n+1}(\d  ).    
\label{f4}
\eeq
In this limit, the CFM are close to NFM  and  neither  
can resolve details of the high multiplicity
tail of  $P_n(\d )$, although both CFM and NFM are known to be 
sensitive to fluctuations of large  multiplicities.  
Indeed, Eq. (\ref{f4}) means $\lan n (\d ) \ran \sim P_1(\d )$ and 
the definition of NFM can be approximated by   
\beq
F_q (\d  ) \simeq  
\frac{ q! P_q(\d ) }{P_1^q(\d  )},
\label{f5}
\eeq
i.e. the NFM are affected by the single-particle probability $P_1(\d )$.  
Thus even if the numbers of particles inside $\d$ follow
a Poisson distribution for large multiplicities, 
$F_q (\d )$  are  not unity 
if the $P_1(\d )$ is not determined by a Poisson value due to some
kinematic reasons unrelated to multiple gluon emissions.

This situation is avoided in the BP which measure the high multiplicities
locally, near $n-1$ multiplicity. (In fact, the BP  are
determined by the second derivative from $\ln P_n(\d )$ at sufficiently
high $n$.) Therefore, even if $P_n (\d )$ for $n<n'$ 
has a non-Poisson behavior, 
this does not affect the property $\eta_q (\d )=1$ for 
events with particle multiplicities $n\geq n'$ 
which obey  the Poisson law. This important property
of the BP will be used in the next section.   

\section{Analytic QCD predictions for the BP}

There exist a number of the analytic QCD predictions for CFM and NFM  
of the parton multiplicity distributions obtained in
the Double Leading Log Approximation (DLLA) \cite{ochs,drem,brax}.
Here we discuss in more details the predictions derived in \cite{ochs},
since the others differ only by different treatment of non-leading
perturbative contributions.

\begin{figure}
\centering
\begin{minipage}[c]{0.4\textwidth}
\centering
\caption{A schematic representation of the measurements of the
probability distribution $P_n(\d, \Theta)$ inside jet.}
\label{jet}
\end{minipage}%
\begin{minipage}[c]{0.6\textwidth}
\centering
\mbox{\epsfig{file=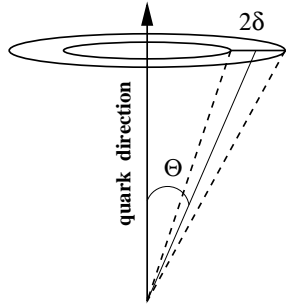,width=0.4\linewidth}}
\end{minipage}
\end{figure}

The angular window considered here is defined for rings 
with the opening angle $\Theta$ around the outgoing-quark direction 
(see Fig.~\ref{jet}). 
The phase-space interval where partons are counted is determined as  
half width $\d$ of the ring. Assuming sufficiently large energies,
the probability $P_n(\d , \Theta)$ to find $n$ partons
inside the interval $2\d$ is dominated by gluon bremsstrahlung off the 
outgoing  quark.       

The analytic QCD predictions can be 
formulated in terms of the variable $z$, which is related to 
$\d$ as  
\beq 
z=a^{-1} \ln (\Theta /\d  ),
\qquad a=\ln (E\Theta / \Lambda ) , 
\label{angdef}
\eeq
where $E$ is the energy of the outgoing quark radiating soft gluons
and $\Lambda$
is the adjustable effective QCD scale. 
A decrease of the angular window $\d$ corresponds
to an increase of the variable $z$. The maximum possible  
phase-space region ($\d = \Theta$) corresponds  to $z=0$. 
The variable $z$ takes the maximum value $z=1$ for $\d = \Lambda /E$.

The CFM derived in the DLLA satisfy the relation   \cite{ochs} 
\beq
\lan n (\d ) \ran^q K_q(z)=
A_q \exp \left[ a (2\>\gamma_0(Q)\> w_q-z) \right],    
\label{an1}
\eeq
\beq
w_q\simeq \sqrt{1-z}\left(q-\frac{\ln(1-z)}{2q}\right)
\label{an2}
\eeq 
with $\gamma_0(Q)=\sqrt{2\,C_{\mathrm{A}}\as(Q)/\pi}$  being  
the anomalous QCD dimension calculated at a scale $Q\simeq  E\Theta$,
$\as$ is the strong coupling constant and $C_{\mathrm{A}}=3$ 
is the gluon color factor.
The functions  $A_q$ are determined by non-leading QCD contributions
and reflect the values  of CFM  in the full  phase space ($z=0$).

It is however clear  from the previous section that the behavior
of CFM at high orders $q$ and large $z$ (small $\d$) is not strictly 
related to the structure
of fluctuations for  multiplicities $n\ge q$. Similar to the NFM,
their values  are affected by the average number $\lan  n (z) \ran$ of 
counted particles which is dominated by $P_1(z)$ for  
sufficiently large  $z$. 
Therefore, to learn more about fluctuations without the bias from  
small-multiplicity events, one should calculate the BP. 
At large $z$, $\lan n (z) \ran^q K_q (z)\simeq q! P_q(z)$ and thus 
the BP are 
\beq
\eta_q(z) \simeq \frac{A_qA_{q-2}}{A^2_{q-1}} \exp 
\left[2 a \gamma_0(Q) (w_q + w_{q-2} - 2 w_{q-1}) \right], 
\label{an3}
\eeq  
and, assuming (\ref{an2}), one finally obtains
\beq
\eta_q(z) \simeq \frac{A_qA_{q-2}}{A^2_{q-1}} \exp
\left[- \frac{2 a \gamma_0(Q) \sqrt{1-z} 
\ln (1-z) }{q (q-1) (q-2)} \right] ,  
\label{an4}
\eeq   
where $q>2$. 
It is remarkable that the expression  in the exponent 
behaves  as $1/q^3$ for large $q$. Therefore, the exponential 
factor for large $q$ is very close to unity, i.e.  
\beq
\eta_q(z) \simeq \eta_q(0)=\frac{A_qA_{q-2}}{A^2_{q-1}}, \qquad q>>1.    
\label{an5}
\eeq
This expression shows that the BP (\ref{an5}) for high orders $q$ 
are close to those measured in the full
phase space at $z=0$, since they are only constructed from the functions
$A_q$ determined by non-leading QCD contributions.
Generally, one expects
the values of BP to be close to unity for $z=0$, if the 
parton multiplicities in the opening angle $\Theta$ 
follow a distribution slightly
broader than a Poisson distribution (i.e. a negative binomial distribution). 
 
A similar behavior of the BP is expected from alternative DLLA  
calculations. For instance, the NFM in the angular windows as defined above 
are given by  \cite{brax} 
\beq
F_q(z)=F_q(0)\exp [z\> a\> (1-D_q)(q-1)], 
\qquad 
D_q\simeq2\,\gamma_0(Q)\frac{q+1}{q}\left(\frac{1-\sqrt{1-z}}{z}\right).
\label{anb1}
\eeq
From (\ref{f5}) and the definition of the BP, one obtains 
\beq
\eta_q(z) \simeq \eta_q(0)\>  \exp
\left[- \frac{4 a \gamma_0(Q) (1-\sqrt{1-z} )}  
{q (q-1) (q-2)} \right] ,   
\label{anb4}
\eeq
\beq
\eta_q(0) = \frac{F_q(0)F_{q-2}(0)}{F^2_{q-1}(0)},  
\label{anb4x}
\eeq
where again the exponential factor in (\ref{anb4}) 
is small at large $q$. Therefore, at large $q$, the BP are close  
to unity even if the angular phase-space region is very small.        
    
The  behavior (\ref{an5}) is opposite to that existing at  
small $z$ (large $\d$). 
It is expected that the BP rise with increase of $z$, 
following a power-like behavior. Such a trend has been found recently
by the L3 Collaboration \cite{fluc_l3} for  
the rapidity distribution averaged over the all available phase space. 
The power-like behavior of the BP is also expected in simple cascade
models \cite{ch}.

\section{Monte Carlo tests}

The direct comparison of the analytic expectation  discussed
above with data is complicated by the fact that the calculation is
performed in the DLLA
which is know to contain a number of  limitations.  
The most significant approximation is neglect of energy-momentum 
conservation  in the gluon splittings (energy recoil effects).
Therefore, it is important to compare the calculations with
the MC parton shower which  includes explicitly the recoil effect.
In  addition, comparisons with the hadron level of MC allow us  
to investigate the
contributions  from hadronization.

To do this, we  used  the ARIADNE Monte
Carlo program \cite{ard} for $e^+e^-$ annihilation into hadrons. 
The hadronic events were generated at 
$\sqrt{s}=500$ GeV with an integrated luminosity
of 300 fb$^{-1}$. The thrust axis was used to determine
the jet direction.  We used  $\Theta=25^0$ with respect to the
jet axis and the effective $\Lambda=0.15$ to
define the variable $z$.     

In addition to the hadronic sample, we generated a
parton sample of MC events after switching off the hadronization phase.
The perturbative phase was  terminated by a cut-off in the transverse
momentum of partons, $Q_0$, which is chosen to be 300 MeV,
to obtain the same average multiplicity of partons inside the 
opening angle $\Theta$ as for hadrons. 
A similar method to analyze the parton correlations has been 
used in \cite{ochs_pt}.  

\begin{figure}[htbp]
\begin{center}
\mbox{\epsfig{file=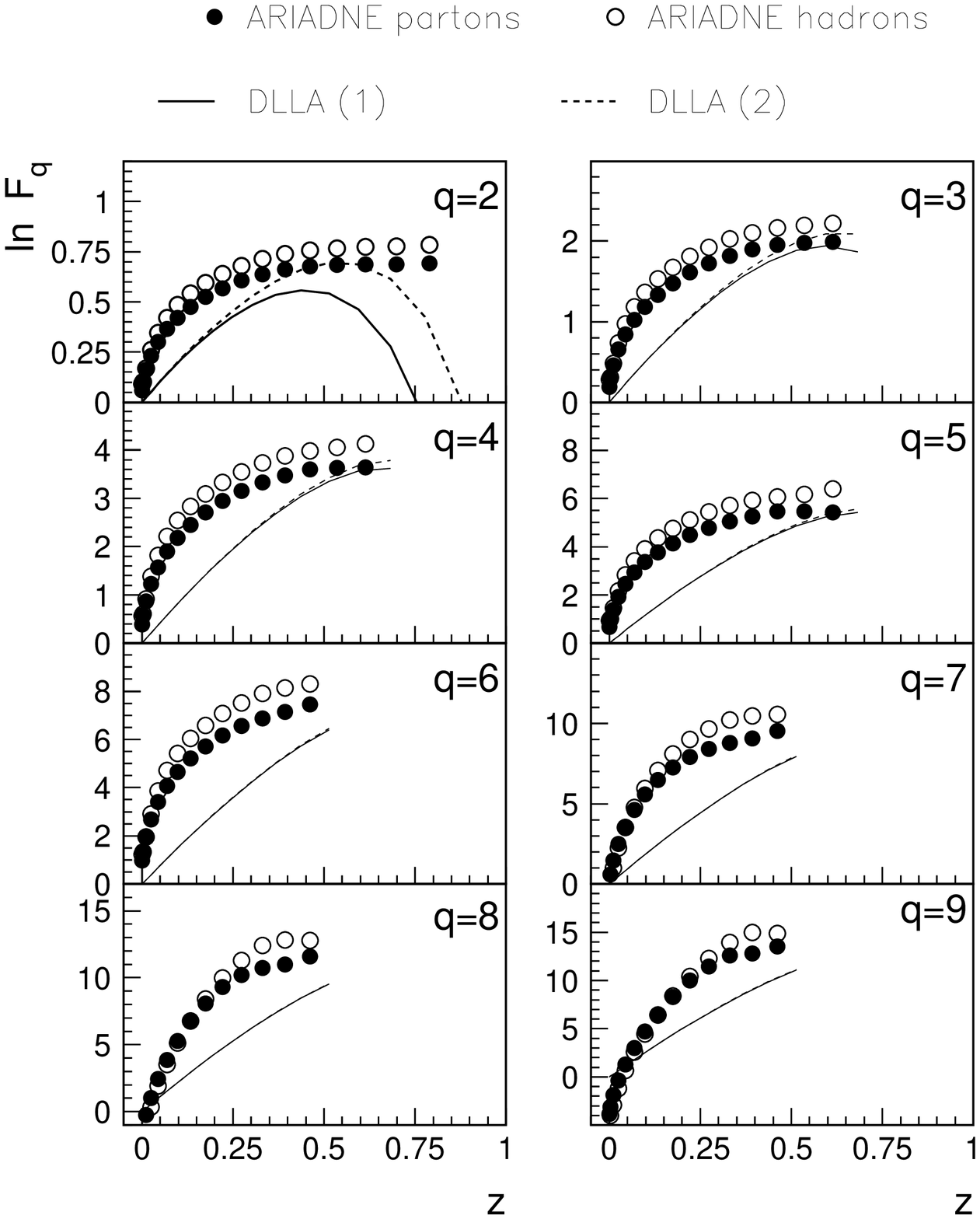,width=0.9\linewidth}}
\end{center}
\caption{
The ARIADNE and analytic DLLA predictions for  
the NFM of different orders as a function of $z$
defined in (\ref{angdef})
for $e^+e^-$ annihilation at 
$\sqrt{s}=500$ GeV$^2$. 
The lines show two different DLLA QCD predictions:  
(1) DLLA \cite{ochs} (solid lines);  (2)
DLLA \cite{brax} (dashed lines).}
\label{p1}
\end{figure}

Figure~\ref{p1} shows the behavior of  NFM for different orders $q$
as a function of $z$, separately for hadrons and partons.  
Such a measurement, for $q=9$,  requires counting at least nine
particles in small angular intervals. 
The smallest size of $\delta$
for the present study is one degree ($z\simeq 0.5$), 
which is close to that used at
LEP and HERA \cite{ang_ex}.

The factorial moments show a saturation at high $z$. The parton-level
predictions systematically overestimate the moments
measured using final-state hadrons. 

In addition to the MC predictions, the NFM
obtained from the analytic DLLA calculations \cite{ochs,brax} are  
also shown in Fig.~\ref{p1}. 
The major difference between these two analytic results is a different
treatment of non-leading QCD contributions to the parton correlations. 
At present, the analytic calculations
cannot give a satisfactory description of the fluctuations,
although they reproduce the rise of the moments. A similar conclusion
has been made   
analysing the experimental data at LEP and HERA \cite{ang_ex}.
For the present comparisons with the MC model, we did not use the
normalization constant $F_q(0)$ (i.e. the NFM 
calculated at $z=0$ or $\Theta = \Theta_0$),
in order to remove the
theoretical uncertainty in the absolute values of $F_q(z)$.
Therefore, the theoretical curves shown in Fig.~\ref{p1} 
have an arbitrary normalization,
which, however, is not relevant  for comparisons of 
the predicted $z$-dependence of the moments. 
   
\begin{figure}[htbp]
\begin{center}
\mbox{\epsfig{file=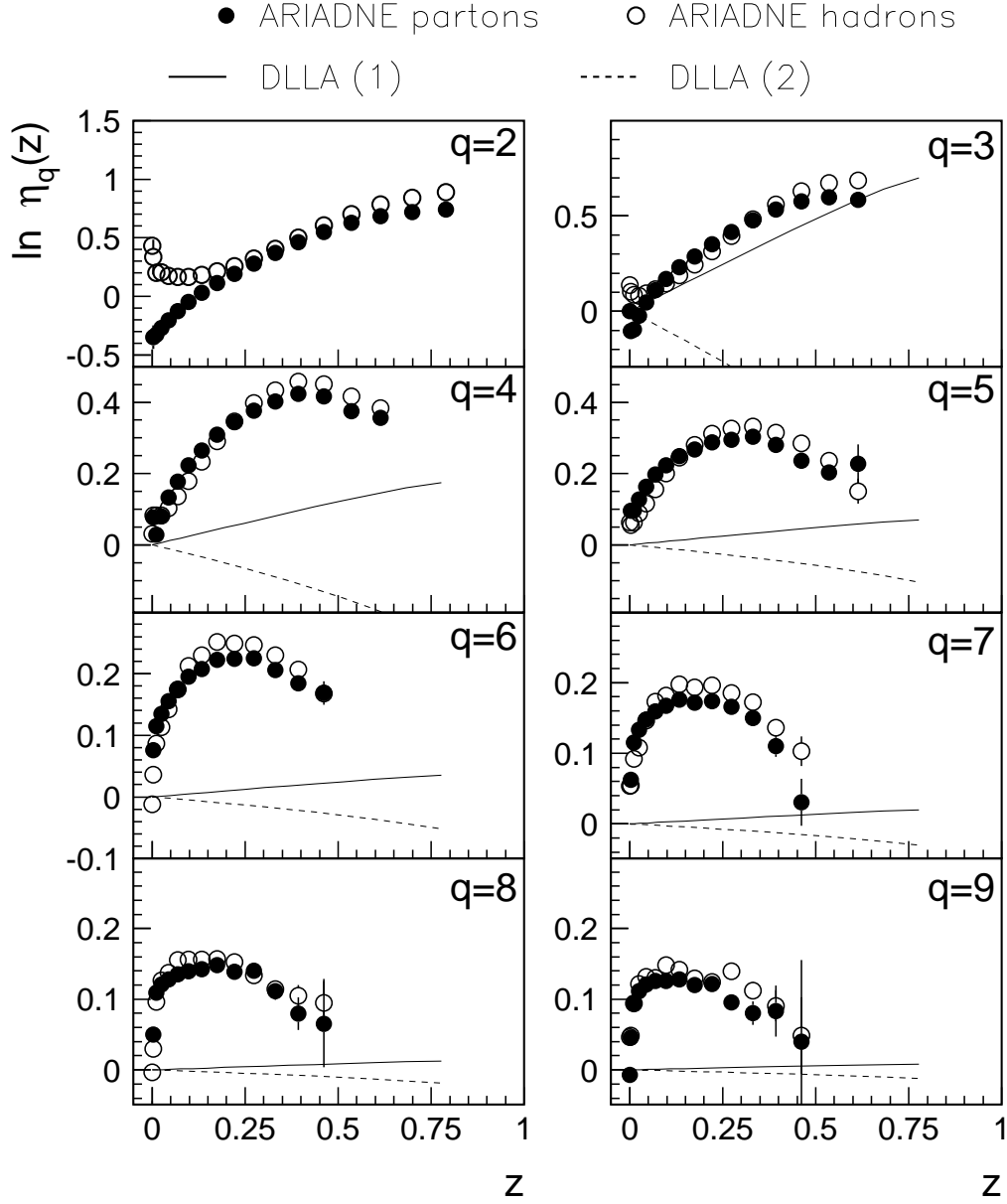,width=0.9\linewidth}}
\end{center}
\caption{
The Monte Carlo and analytic DLLA predictions for  
the BP  of different orders as a function of $z$
defined in (\ref{angdef})
for $e^+e^-$ annihilation at
$\sqrt{s}=500$ GeV$^2$. The lines show the BP  from
the analytic considerations:  
(1) DLLA of Eq.~(\ref{an4}) (solid lines);  (2)
DLLA of Eq.~(\ref{anb4}) (dashed lines). 
Due to the approximation used  to derive the BP,
the analytic curves cannot be considered as a reliable prediction
for small $z$.  
}   
\label{p2}
\end{figure}

Figure~\ref{p2} shows the BP as a function of $z$ for ARIADNE,
separately for hadrons and for partons. The
analytic estimates (\ref{an4}) and (\ref{anb4}) of the BP
for $q>2$ are also shown. The theoretical results can  
only be considered to be a realistic 
for large $z$ values, because of the approximation $P_n(z)\gg P_{n+1}(z)$
used to derive the BP from the CFM or NFM.
It is important to note also that the analytic results 
(\ref{an4}) and (\ref{anb4}) contain  
an  unknown\footnote{
It is possible to consider  
$\ln (\eta_q(z)/\eta_q(0))$, in order to  remove 
the theoretical uncertainty related to the full 
phase-space regions, similar to  $\ln (F_q(z)/F_q(0))$ 
used in  \cite{ang_ex}. For the BP, however, 
this leads to an unstable result due to different
values of the BP at $z=0$ for partons and hadrons. Therefore,
such a normalization was not used in Figs.~\ref{p1} and \ref{p2}. }    
factor $\eta_q(0)$ (\ref{anb4x}), which is expected to be very close to unity.
Therefore, calculating the analytic curves shown in  Fig.~\ref{p2},
a contribution of $\ln \eta_q(0)$ is neglected.  

\vspace{0.3cm}

From this study, the following observations can be made: 

\begin{itemize}
\item
High order BP for partons at large $z$ show a clear tendency to approach 
the values close to those seen at $z=0$.  
This is consistent with our analytic estimates in DLLA,
suggesting a suppression of high-multiplicity 
fluctuations in small phase-space windows.

\item
The hadron-level predictions show the same trend as partons. The values
of BP for hadrons are closer
to those for partons than for the NFM. 
This illustrates relative insensitivity
of the BP to hadronization effects. 
The largest discrepancy between
parton and hadron levels is seen at low $z$  
for $\eta_2 (z)$ which is affected by
the probability $P_0(z )$ to find an event
without particles. 

\item
For the analytic QCD calculations,
there is a large difference
between the two analytic DLLA QCD results, 
especially for the lowest order of BP.
At large $q$ and $z$, the analytic results and the MC predictions
approach a common value ($\ln \eta_q(z)\simeq 0$).    
      
\end{itemize}
 
Thus we have established a very important observation which can
shed light on the jet structure at small scales.
According to the DLLA analytic
calculations and our MC simulations, 
fluctuations of partons  
for sufficiently small angular windows are depleted. 

This study also illustrates that even for the relatively modest c.m. energy
and luminosity, the measurement of the NFM and BP can be possible  
up to $q=9$, thus a LC will allow to observe the effect. 
So far the BP have been
measured in $e^+e^-$ annihilation up to $q=5$, but the 
suppression of multi-gluon fluctuations in small rapidity regions  
has not been observed \cite{fluc_l3}. 

\section{High-multiplicity fluctuations in jets} 

In this section we will summarize the 
observations made in this papers: 

The multiplicity fluctuations of particles for 
large phase-space regions ($z\sim 0$) are rather small,
$F_q(z=0)\sim \eta_q(z=0)\sim 1$. Note that the fluctuations
do not exhibit the Poisson structure since the BP and NFM show   
deviations of the NFM and BP from unity. 

With decrease of the angular window $\d$, the fluctuations  
start to emerge. At medium size of the window, $z\simeq 0.2-0.4$,
the fluctuations are maximal.   

For very small phase-space regions, the fluctuations are depleted  
and approach the values similar to those seen in the full phase space.
This observation is only possible after  the use
of BP which are not affected by low-multiplicity part of
the probability distributions, in contrast to the NFM and CFM. 

This peculiar feature of the fluctuations at low $\d$ can be understood
in terms of strong restriction of the available
phase space for the next generations of partons in the cascade.
One possible explanation of this saturation is to assume 
that particles counted inside $\d$ are produced 
by different parent partons and thus they are   
quasi-independent.  An experimental verification
of this phenomenon is rather important as it 
reveals the dynamics of the parton-shower system and can be 
directly related to the minimum transverse momenta $Q_0$ used 
in the perturbative calculations.       

The expected suppression of high-multiplicity fluctuations,   
however, seems to be different to that obtained in Refs. \cite{ochs_pt,ochs},  
which also suggest a reduction of fluctuations  in  restricted 
$p_t$ intervals, or the existence of a critical angle $\d^{crit}$ beyond
which the parton correlations are small and affected by the QCD cut-off. 
The latter effects  should even be seen for the lowest order 
correlation functions and CFM, while the saturation discussed here
is only expected in the tail of the multiplicity distribution
measured with the BP.    

The fact that high-order BP are not sensitive to the hadronization
dynamics  for small $\d$ suggests  applicability of  
the hypothesis of Local Parton-Hadron  Duality \cite{AZIM}
to fluctuations of  high-multiplicity events. 
Therefore, the BP are the most
suitable tool to study the high-multiplicity fluctuations
and comparing experimental  data with perturbative QCD predictions
for partons.  
Some significant differences between fluctuations of partons and hadrons
exist only for $\eta_2$, i.e. for low-multiplicity events.

{}

\newpage

\newpage

\end{document}